\documentclass[letterpaper,floatfix,aps,prl,amsmath,twocolumn,
superscriptaddress,showpacs,
nofootinbib]{revtex4}

\usepackage{epsfig}
\usepackage{verbatim}
\usepackage{epsf}
\usepackage{array}
\usepackage{hyperref}

\newcommand{\be}{\begin{equation}}
\newcommand{\ee}{\end{equation}}
\newcommand{\bea}{\begin{eqnarray}}
\newcommand{\eea}{\end{eqnarray}}

\renewcommand{\Re}{\mathrm{Re}\,}
\renewcommand{\Im}{\mathrm{Im}\,}

\newcommand{\eref}[1]{Eq.~(\ref{#1})}
\newcommand{\esref}[1]{Eqs.~(\ref{#1})}
\newcommand{\rref}[1]{(\ref{#1})}

\newcommand{\ocite}[1]{Ref.~\onlinecite{#1}}

\newcommand{\qp}{\mathrm{qp}}

\newcommand{\EZO}{\omega_{10}}
\newcommand{\ka}{x_{\mathrm{qp}}}
\newcommand{\kaA}{x^{\mathrm{A}}_{\mathrm{qp}}}
\newcommand{\Gif}{\Gamma_{i\to f}}
\newcommand{\Goz}{\Gamma_{1\to 0}}

\begin{document}

\title{Quasiparticle relaxation of superconducting qubits in the presence of flux}

\author{G. Catelani}
\affiliation{Departments of Physics and Applied Physics, Yale
University, New Haven, CT 06520, USA}
\author{J. Koch}
\affiliation{Department of Physics and Astronomy, Northwestern University, Evanston, IL 60208, USA}
\author{L. Frunzio}
\author{R. J. Schoelkopf}
\author{M. H. Devoret}
\author{L. I. Glazman}
\affiliation{Departments of Physics and Applied Physics, Yale
University, New Haven, CT 06520, USA}

\begin{abstract}
  Quasiparticle tunneling across a Josephson junction sets a limit for
  the lifetime of a superconducting qubit state. We develop a general
  theory of the corresponding decay rate in a qubit controlled by a
  magnetic flux. The flux affects quasiparticles tunneling amplitudes,
  thus making the decay rate flux-dependent. The theory is applicable
  for an arbitrary quasiparticle distribution. It provides estimates
  for the rates in practically important quantum circuits and also
  offers a new way of measuring the phase-dependent admittance of
  a Josephson junction.
\end{abstract}

\date{\today}


\maketitle

Long coherence times of superconducting qubits rely
on the decoupling of the order parameter quantum oscillations
from other low-energy degrees of freedom.
Quasiparticles provide an intrinsic set of states capable of
exchanging energy with the qubit degree of freedom. In equilibrium,
quasiparticle populations should get completely depleted at low temperatures, rendering
the corresponding qubit relaxation channel ineffective. In practice,
however, some nonequilibrium quasiparticles
are observed~\cite{schreier,Martinis}.
The question whether the relaxation driven by
quasiparticles is comparable to the extrinsic mechanisms of qubit
relaxation has remained open.

The theory of quasiparticle relaxation was addressed in \cite{lutchyn1} for
a charge qubit, whose computational space consists of two values of
charge of the same parity (even or odd) stored in a Cooper pair box. The elementary
process of relaxation essentially amounts to the well-studied
quasiparticle poisoning \cite{matveev,joyez}: a quasiparticle entering the
Cooper pair box changes the parity of the state. Later, the theory
\cite{lutchyn1} was modified to estimate the effect of quasiparticles in a
``transmon'' device \cite{transmon}, where the dominant energy scale comes
from the Josephson inductance.
Quantum
fluctuations of phase in a transmon are relatively small, while
the uncertainty of charge in the qubit states is significant. The
advantage of the transmon is its low sensitivity to charge
noise. Further consideration of the relaxation induced by
quasiparticles in superconducting qubits was developed in \ocite{Martinis}.
The properties of qubits such as the phase and flux qubits
\cite{Dev_rev}, the transmon,
and the newly developed fluxonium \cite{flux_exp}
can be tuned by threading a
magnetic flux through the device.

Here, we predict that the relaxation rates induced by quasiparticle
tunneling depend on the magnetic flux threading the qubit.
The theory we develop
allows for any quasiparticle distribution and
any magnitude of quantum fluctuations of the
phase of the order parameter; thus, it provides relaxation rates for
the entire spectrum of superconducting qubits.  We also show that a
spectroscopic measurement on a device designed to have small phase
fluctuations may enable one to measure the enigmatic phase-dependence
of the Josephson junction
admittance~\cite{BP}.

\begin{figure}[b]
\begin{center}
\includegraphics[width=0.36\textwidth]{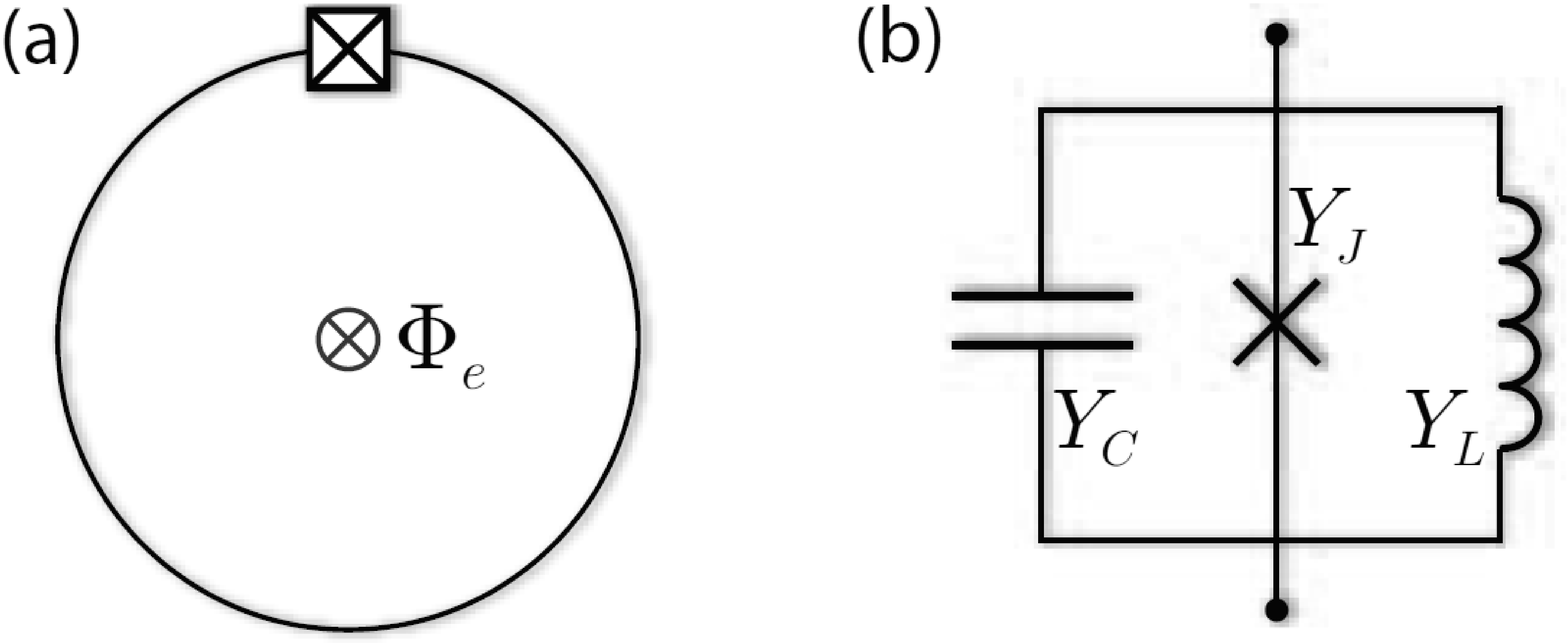}
\end{center}
\caption{(a) Schematic representation of a qubit controlled by a magnetic flux,
see \eref{Hphi}.
(b) Effective circuit diagram with three parallel elements
-- capacitor, Josephson junction, and inductor -- characterized by their respective admittances.}
\label{fig}
\end{figure}

We consider the system consisting of a Josephson junction
closed by an inductive loop, see Fig.~\ref{fig}.  The Hamiltonian $\hat{H}$
governing the low-energy dynamics of the system can be divided into
three parts
\be\label{Htot}
\hat{H} = \hat{H}_{\varphi} + \hat{H}_{\qp} + \hat{H}_T \, .
\ee
The first term takes the form of the inductively shunted Josephson
junction
\be\label{Hphi}
\!\!\hat{H}_{\varphi}\! = 4E_C \!\left(\!\hat{N}-n_g\!\right)^2 \!\!-E_J \cos \hat\varphi
+ \frac12 E_L\!\left(\hat\varphi-2\pi\frac{\Phi_e}{\Phi_0}\right)^2,
\ee
where
$\hat{N}=-id/d\varphi$ is the number operator of Cooper
pairs passed across the junction, $n_g$ is the dimensionless gate voltage, $\Phi_e$ is the external flux
threading the loop, and $\Phi_0=h/2e$ is the flux quantum.
With appropriate choices for the parameters characterizing the qubit -- charging energy $E_C$,
Josephson energy $E_J$, and inductive energy $E_L$ -- and of the biases $n_g$ and $\Phi_e$, \eref{Hphi}
can describe a single-junction qubit and also a multi-junction one, so long as
an array of junctions with energies $E_J^a \gg E_J$ is treated as
an effective inductance~\cite{effind}.

The quasiparticle term $H_\qp$ is the sum of the BCS
quasiparticle Hamiltonians for the left and right leads
\be
\hat{H}_{\qp} = \sum_{j=L,R} \hat{H}_{\qp}^j \, , \quad \hat{H}_{\qp}^j = \sum_{n,\sigma} \epsilon_{n}^j
\hat\alpha^{j\dagger}_{n\sigma} \hat\alpha^j_{n\sigma}.
\ee
Here $\hat\alpha^j_{n\sigma}$($\hat\alpha^{j\dagger}_{n\sigma}$) is the quasiparticle annihilation (creation)
operator, $\sigma=\uparrow, \downarrow$ accounts for spin, and the
quasiparticle energies are $\epsilon^j_{n} =
\sqrt{(\xi_{n}^j)^2+(\Delta^j)^2}$, with $\xi_{n}^j$ and $\Delta^j$ being the
single-particle energy level $n$ in the normal state of lead $j$, and
the gap parameter in that lead, respectively.  Finally, the
tunneling term $\hat{H}_T$
describes quasiparticle tunneling across the junction,
\bea\label{HT}
\hat{H}_{T} = t\!\!\sum_{n,m,\sigma}\!\!
\left(e^{i\frac{\hat\varphi}{2}} u_{n}^L u_{m}^R
- e^{-i\frac{\hat\varphi}{2}} v_{m}^R v_{n}^L
\right)\hat\alpha_{n\sigma}^{L\dagger} \hat\alpha^R_{m\sigma}
+ \text{H.c.}\quad
\eea
The electron tunneling amplitude $t$ here is related to the junction conductance
$g = 4\pi e^2 \nu^L \nu^R t^2/\hbar$.
We work in the tunneling limit $t\ll 1$ and assume identical
densities of states per spin direction in the leads,
$\nu^L\!=\!\nu^R\!=\!\nu_0$.
The Bogoliubov amplitudes $u^j_{n}$, $v^j_{n}$
can be taken real, since \eref{HT} already accounts explicitly for
the phases of the order parameters in the leads
via the gauge-invariant phase difference~\cite{BP} in the exponentials.
Accounting for Josephson effect and quasiparticles dynamics by
Eqs.~(\ref{Hphi})-(\ref{HT}) is possible as long as
the qubit energies and characteristic energies of quasiparticles
(measured from $\Delta$) are small compared to $\Delta$~\cite{lutchyn1}. In this
low-energy limit, we may approximate $u^j_{m} \simeq v^j_{n} \simeq
1/\sqrt{2}$.
Then the operators $e^{\pm i\hat\varphi/2}$ in \eref{HT}
which describe transfer of charge $\pm e$ across the junction,
combine to $\sin (\hat\varphi/2)$. The superposition of the tunneling
amplitudes containing $e^{i\hat\varphi/2}$ and $e^{-i\hat\varphi/2}$
is a manifestation of interference between tunneling of particle-
and hole-like excitations, possible in the presence of the Cooper
pair condensate.
We stress that the phase $\hat\varphi$ is an operator subject to quantum fluctuations, as determined
by $\hat{H}_\varphi$, not an externally controlled parameter
as in the ``classical'' Josephson junction~\cite{BP}.
Moreover, the non-linear nature of phase-quasiparticle coupling is
in stark contrast with the linear coupling between phase and
the electromagnetic environment.

In general, the qubit Hamiltonian in \eref{Hphi} has a discrete
low-energy spectrum.  The tunneling term, \eref{HT}, couples the qubit
to the quasiparticles; therefore, a transition between initial,
$|i\rangle$, and final, $|f\rangle$, qubit states becomes possible
when a quasiparticle is excited during a tunneling event.
The transition rate $\Gif$ can be calculated
using Fermi's Golden Rule,
\be\begin{split}
\Gif = \frac{2\pi}{\hbar} \sum_{\{\lambda\}_\qp} &
\langle\!\langle  \left|\langle f , \{\lambda\}_\qp | \hat{H}_T |i, \{\eta\}_\qp \rangle\right|^2
 \\ & \ \times
 \delta\left(E_{\lambda,\qp}-E_{\eta,\qp}-\hbar\omega_{if}\right) \rangle\!\rangle_\qp \,,
\end{split}\ee
where $\hbar\omega_{if}$ is the energy difference between the qubit states,
$E_{\eta,\qp}$ and $E_{\lambda,\qp}$ are the total energies of the
quasiparticles in their respective initial $\{\eta\}_\qp$ and final
$\{\lambda\}_\qp$ states; double angular brackets
$\langle\!\langle \ldots \rangle\!\rangle_\qp$ denote averaging over the
initial quasiparticle states whose occupation is determined by the distribution functions
$f^j(\xi^j_{n})= \langle\!\langle \hat\alpha_{n\uparrow}^{j\dagger} \hat\alpha^j_{n\uparrow}\rangle\!\rangle_\qp
= \langle\!\langle \hat\alpha_{n\downarrow}^{j\dagger} \hat\alpha^j_{n\downarrow}\rangle\!\rangle_\qp$
($j=L,R$) assumed to be independent of spin.
In terms of the matrix elements of the operator $\sin (\hat\varphi/2)$ we find
\be\label{wif_gen}
\Gif = \left|\langle f|\sin \frac{\hat\varphi}{2}|i\rangle\right|^2
S_\qp\left(\omega_{if}\right).
\ee
In deriving \eref{wif_gen}
we have made a simplifying assumption of equal
gaps, $\Delta_L=\Delta_R=\Delta$.
More importantly,
we concentrate on the case of low-lying excitations,
assuming the characteristic energy $\delta E$ of the quasiparticles
(determined by the distribution functions; in thermal equilibrium, $\delta E=k_B T$) and
the energy of the qubit transition
$\hbar\omega_{if}$ are small, $\hbar\omega_{if}\,,\delta E \ll
2\Delta$.
That enables us to
factorize the transition rate $\Gif$ into a product of terms
which depend separately on the qubit dynamics and quasiparticle kinetics.
The latter determines the
normalized quasiparticle current spectral density
$S_\qp$,
\bea\label{tF_def}
  S_\qp\left(\omega\right) & = &
  \frac{8E_J}{\pi\hbar}
  \int_0^{\infty}\!\!\!dx\,
  \frac{1}{\sqrt{x}\sqrt{x+\hbar\omega/\Delta}}\Big[ f_E^L \left((1+x)\Delta\right) \nonumber \\
  & \times & \left(1-f_E^R\left((1+x)\Delta+\hbar\omega\right)\right) + (L
  \leftrightarrow R) \Big],
\eea
where $\omega>0$~\cite{negom} and we used the relation $E_J = g\Delta/8g_K$;
$g_K = e^2/h$ is the conductance quantum.
The integrand in $S_\qp$ equals, up to a
factor, the rate of transitions of a quasiparticle between the initial
and final states, properly weighted with their occupation probabilities
(their energies are $(1+x)\Delta$ and $(1+x)\Delta+\hbar\omega$,
respectively). These probabilities are expressed in terms of the
quasiparticle energy distribution functions,
$f_E^{j} (\epsilon) =\left(f^{j}(\xi) + f^{j}(-\xi)\right)/2$, with
$\epsilon = \sqrt{\xi^2+\Delta^2}$ and $j=L,R$.

We notice that $S_\qp$ is related to the real part of the
quasiparticle contribution to the ``classical''~\cite{BP}
Josephson junction admittance at zero phase difference,
\begin{equation}
\label{SY}
S_\qp(\omega)-S_\qp(-\omega) = \frac{\omega}{\pi} \frac{1}{g_K}\Re
Y_\qp (\omega),
\end{equation}
at arbitrary ratio $\hbar\omega/\delta E$ and for any
quasiparticle distribution function. In the rest of this Letter we
consider the ``high-frequency'' limit~\cite{footnote3}
$\hbar\omega/\delta E \gg 1$ (but still $\hbar\omega\ll
2\Delta$),
and equal populations on the two sides of the junction, $f^L = f^R$.
Then we can simplify Eq.~(\ref{SY}),
\be\label{S_Y}
S_\qp(\omega) = \frac{\omega}{\pi} \frac{1}{g_K}\Re Y_\qp (\omega)\,,
\qquad \omega>0 \, ,
\ee
and express $\Re Y_\qp$ in terms of the
quasiparticle density
\be\label{nqp}
n_\qp = 2\sqrt{2}\nu_0\Delta \int_0^\infty\frac{dx}{\sqrt{x}}
f_E((1+x)\Delta)
\ee
(written using the same approximations as above) as
\be\label{tF_hf}
\Re Y_\qp(\omega)  =  \frac12 \ka g
\left(\frac{2\Delta}{\hbar\omega}\right)^{3/2} \, ,
\quad \ka = \frac{n_{\qp}}{2\nu_0 \Delta}.
\ee
Here $\ka$ is the quasiparticle density normalized by the density of
Cooper pairs; in thermal equilibrium, $\ka=\sqrt{2\pi k_B T/\Delta}\, e^{-\Delta/k_B T}$.
Under the above stated assumptions, \eref{tF_hf}
applies to an otherwise arbitrary distribution function, and hence to
nonequilibrium conditions.

The structure of $S_\qp$ is identical to the corresponding one in
the Mattis-Bardeen formula~\cite{mb}, when generalized to the
case of non-equilibrium distributions~\cite{cng}. The
connection
between the Mattis-Bardeen theory, which describes absorption of
electromagnetic waves impinging on the surface of a ``dirty''
superconductor, and the theory of qubit relaxation caused by
quasiparticles in a Josephson junction was
noticed in~\cite{Martinis}.
At the  microscopic level, dissipation in the two settings is caused by
quasiparticle transitions occurring without momentum conservation
(both systems lack translational invariance). The difference
between the two problems is in the form of the perturbation causing the
transitions. In our case, the quasiparticle transitions are
due to the coupling to
the qubit degree of freedom.
The matrix element $\langle f|\sin(\hat{\varphi}/2)|i\rangle$ in Eq.~(\ref{wif_gen})
characterizes that coupling and it is sensitive to flux~\cite{footnote}.

We stress that
\eref{wif_gen} holds for any single-junction qubit, its properties being encoded in
the wavefunctions $|i\rangle$, $|f\rangle$ entering the matrix
element.
For a qubit
comprising multiple junctions
the tunneling Hamiltonian \eref{HT} and hence
\eref{wif_gen} are
given by a sum over junctions, with $\hat\varphi$ being replaced by the
phase difference across each junction.
In a multi-junction qubit, the states $|i\rangle$ and $|f\rangle$ depend on
all the phase differences (up to the constraint set by fluxoid quantization~\cite{orlando}).

We focus first on the case of a weakly anharmonic system, which
already reveals a non-trivial flux dependence of relaxation.  Its
low-lying states, as the example of the transmon
shows~\cite{transmon}, can be used as qubit states. We
assume $E_L\neq 0$ to eliminate $n_g$ by a gauge
transformation~\cite{flux_th}.
In the transmon $E_L=0$, but the $n_g$-dependent corrections are exponentially small~\cite{transmon};
hence, the results below can be applied to a single-junction transmon
by setting $\varphi_0=0$.
In the limit $E_C \ll E_J, E_L$,
\eref{Hphi} describes an ideal LC circuit with a junction in parallel.
The phase across the junction is determined, up to small fluctuations,
by the external magnetic flux: by minimizing the potential energy part
of \eref{Hphi}, we find that the phase $\varphi_0$ at the minimum
satisfies
\be\label{vp0_Pe}
E_J \sin\varphi_0 + E_L(\varphi_0 - 2\pi\Phi_e/\Phi_0) =0.
\ee
Near this minimum, the system behaves as a weakly anharmonic
oscillator.  Anharmonicity and quality factor $Q$ determine the
operability of the system as a qubit~\cite{Dev_rev}.  In the present
case, the operability condition can be written as $Q/n_w \gg
1$, where $n_w$ is the number of levels in the anharmonic well.
For large $Q$ the system can therefore be operated as a qubit despite
its weak anharmonicity.  To the leading order in $E_C/(E_L + E_J \cos
\varphi_0)$~\cite{footnote2} and for low-lying levels $n \ll n_w$,
we can neglect
anharmonic corrections when evaluating the matrix element in
\eref{wif_gen}~\cite{phasequ}. Using standard results for the
harmonic oscillator, its value for transitions between two neighboring
levels is approximated as:
\be
\left|\langle n-1|\sin \frac{\hat\varphi}{2}|n\rangle \right|^2 =
n
\frac{E_C}{\hbar\EZO}
\frac{1 + \cos \varphi_0}{2},
\label{harmonic}
\ee
with $\EZO=\sqrt{8E_C (E_L + E_J \cos \varphi_0)}/\hbar$.  In deriving
Eq.~(\ref{harmonic}), we replaced $\sin(\hat{\varphi}/2)$ by its
linear expansion, which sets the limit on the excitation level, $n \ll
\hbar\EZO/E_C$. For higher $n$ quantum fluctuations of phase are
significant and transitions between distant levels proliferate, due to
the nonlinearity of the operator $\sin(\hat{\varphi}/2)$.

Concentrating on $n\ll\hbar\EZO/E_C$, we use $E_C=e^2/2C$ and Eqs.~(\ref{wif_gen}),
(\ref{S_Y}), and (\ref{harmonic}) to find
\be\label{wnn}
\Gamma_{n\to n-1}  =  \frac{n}{C} \Re Y_\qp(\EZO)\frac{1+\cos\varphi_0}{2}\, .
\ee
For the transition between the two lowest states ($n=1 \to 0$) and in
the absence of magnetic flux ($\varphi_0=0$), \eref{wnn} reduces to
the expression for the rate of \ocite{Martinis}.

Using now Eq.~(\ref{wnn}) with $n=1$, we find
the inverse of the transition
$Q$-factor at arbitrary $\varphi_0$,
\be\label{tand1}
\frac{1}{Q_{10}} = \frac{\Goz}{\EZO} =
\frac{1}{\pi g_K}\Re Y_\qp(\EZO) \frac{E_C}{\hbar \EZO}
\frac{1+\cos \varphi_0}{2},
\ee
Note that $\varphi_0$ and hence the transition frequency $\EZO$ depend
on the external flux $\Phi_e$, see \eref{vp0_Pe}.
We stress that the flux dependence
presented here is specific to the single-junction case;
results for multiple-junction qubits obtained with the developed method will be presented elsewhere.
In the limit $E_J \gg E_L$, the flux dependence can be neglected and
the last factor on the right hand side of \eref{tand1} reduces to $\approx 1$.
Qubit $Q$-factors measured at temperatures $\sim 20$~mK are in the range
$10^4$-$10^5$~\cite{Houk}.
Using typical parameters for Al-based qubits ($\Delta \sim 2\times 10^{-4}$~eV,
$\omega_{10}/2\pi \sim 1-10$~GHz) we find that to reproduce the
experimental $Q$-factors with \eref{tand1} and a thermal equilibrium-like quasiparticle density,
we must assume an effective quasiparticle temperature an order
of magnitude larger than the base temperature.
This points either to the quasiparticles being
in a non-equilibrium state whose origin is at present unclear,
or to the prevalence of extrinsic relaxation mechanisms.

Beside causing dissipation, quasiparticle tunneling leads to a shift
in the resonant frequency of the circuit.  In the regime under
consideration, the resonant frequency is the zero of the total
admittance $Y(\omega)$ which for the parallel elements representing
the qubit (Fig.~\ref{fig}b) is $Y=Y_C+Y_L+Y_J$, with $Y_C=i\omega C$,
$Y_L=1/i\omega L$, and the junction admittance $Y_J$ being the
sum~\cite{BP} of a purely inductive Josephson term and a quasiparticle
term,
\be
Y_J(\omega,\varphi) = \frac{(1-2\kaA)}{iL_J\omega}\cos\varphi + Y_{\qp}(\omega)
\frac{1+\cos\varphi}{2}\,.
\ee
Here $L_J=\hbar/\pi g\Delta$ is the conventionally defined Josephson
junction inductance, $\kaA=f_E(\Delta)$ has the meaning
of the population of Andreev levels~\cite{beenakker} ($\kaA = e^{-\Delta/k_B T}$ in thermal
equilibrium), and
\be\label{Yqp}
Y_{\qp}(\omega) =
  -\frac{2}{iL_J\omega}
  \left[\frac{\ka}{\pi}\sqrt{\frac{\Delta}{i \hbar\omega}}
        - \kaA \right]
\ee
is the quasiparticle admittance at zero phase difference
in the high-frequency limit [its real part agrees with \eref{tF_hf}].
Both free ($\ka$) and bound ($\kaA$) quasiparticles affect $Y_J$.
The frequency shift they cause, measured from
the frequency in the absence of quasiparticles, is found by solving
$Y_C+Y_L+Y_J =0$ at linear order in $\ka$, $\kaA$:
\be
\delta\omega =
\frac{i}{2C} Y_\qp (\EZO)
\frac{1+\cos\varphi_0}{2}
 -\frac{\pi g \Delta}{C
  \hbar\EZO}\kaA\cos\varphi_0\,.
\label{deltaomega}
\ee
The imaginary part of frequency shift reproduces -- when \eref{Yqp} for the admittance is used
-- half the dissipation rate $\Goz$ calculated above, while the real part
in the high-frequency limit equals
\bea\label{freq_shift}
\Re \delta\omega =
  \frac{1}{2} \frac{\omega_p^2}{\EZO}
  &&\!\bigg[
  \kaA\left(1-\cos\varphi_0\right)
  \\ && - \frac{\ka}{2\pi}
 \sqrt{\frac{2\Delta}{\hbar\EZO}}\left(1+\cos\varphi_0\right)
      \,\bigg], \nonumber
\eea
where $\omega_p = \sqrt{8E_JE_C}/\hbar$ is the junction plasma frequency.
The relation Eq.~(\ref{deltaomega}) for $\delta\omega$ in the
weakly anharmonic regime
opens new ways
to study experimentally the effect of quasiparticles on the
Josephson junction admittance.
It may elucidate the phase dependence
of its dissipative part, where experimental data are still
controversial.
Measuring $\Im\delta\omega$ along with $\Re\delta\omega$, \esref{deltaomega}-\rref{freq_shift},
may also shed light on the nature of nonequilibrium quasiparticle distributions:
unlike the dissipative part, $\Re\delta\omega$ depends on both $\kaA$
and $\ka$.
Experimental efforts to verify the flux and quasiparticle density dependence of
$\delta\omega$
are in progress in our group and elsewhere~\cite{martinis1}.

The dependence of relaxation on flux is also sensitive to the states
involved in the transition.  The interaction between phase and
quasiparticles is non-linear (the form of the non-linearity is
associated with the discreteness of charge, as discussed above), thus
allowing transitions between distant levels even in a
harmonic-oscillator potential.  At $E_C/\hbar\omega_{10} \ll 1$ these
transitions are suppressed by the smallness of quantum
fluctuations of the phase.  For example, $\Gamma_{2\to 0}$ appears
only in the second order in fluctuations of $\hat\varphi$ around $\varphi_0$,
\be\label{G2}
\!\!\Gamma_{2\to 0} =
\frac{2\EZO}{\pi} \frac{1}{g_K} \Re Y_\qp(2\EZO)\!
\left(\frac{E_C}{\hbar \omega_{10}}\right)^2\!
 \frac{1-\cos\varphi_0}{4}.
\ee
Here $\Re Y_\qp$ is given by \eref{tF_hf}. Notice the difference in
the $\varphi_0$ dependence between Eqs.~(\ref{wnn}) and (\ref{G2}).

The effect of nonlinear in $\hat\varphi$ coupling of the qubit degree
of freedom to quasiparticles is striking for a system with $E_J > E_L$ biased near half
the flux quantum~\cite{Zorin}.
In the case of a flux qubit, the potential has a pronounced
double-well shape and the qubit states $|0\rangle$ and
$|1\rangle$ are the lowest tunnel-split eigenstates in this
potential~\cite{Dev_rev}.
The rate $\Gamma_{1\to 0}$ vanishes at
$\Phi_e=\Phi_0/2$ due to the destructive interference: at that point,
the potential in Eq.~(\ref{Hphi}) and function $\sin \varphi/2$ in
Eq.~(\ref{wif_gen}) are symmetric around $\varphi=\pi$, while the
qubit states $|0\rangle$, $|1\rangle$ are symmetric and antisymmetric,
respectively. (The symmetry and its consequences are missed if the
$\sin \hat\varphi/2$ interaction is replaced with the linear
phase-quasiparticle coupling accepted in the environmental approach.)
To evaluate $\Gamma_{1\to 0}$ at finite $\Phi_e-\Phi_0/2$ for a qubit governed by \eref{Hphi}
-- {\sl i.e.,} a single junction connected to an (effective) inductor --
we consider the case of tunnel splitting~\cite{transmon,flux_th}
\be
\epsilon_0 = 4\sqrt{\frac{2}{\pi}} \, \hbar\omega_p \left(\frac{8E_J}{E_C}\right)^{1/4}
e^{-\sqrt{8E_J/E_C}}
\ee
small compared to inductive and plasma
energies, $\hbar\omega_p \gg E_L \gg \epsilon_0$.
Then we can evaluate the matrix
element in \eref{wif_gen} in a tight-binding approximation, with
the qubit states given by linear combinations of states localized in the two wells. Using
\eref{S_Y}, we arrive at
\be\label{wif_fq}
\Gamma_{1 \to 0} =
\frac{\EZO}{\pi}\frac{1}{g_K} \Re Y_\qp(\EZO)
\left(\frac{\epsilon_0}{4\pi E_J}\right)^2
\left(1 - \frac{\epsilon_0^2}{(\hbar\EZO)^2}\right),
\ee
where the transition frequency is related to the flux by $\EZO =
\sqrt{\epsilon_0^2 + \left[(2\pi)^2E_L (\Phi_e/\Phi_0-1/2)\right]^2}/\hbar$
and $\Re Y_\qp$ is given in \eref{tF_hf}.

Losses in other elements of the qubit give additional contributions to the
relaxation rate in \eref{wif_fq}.
For example, if the qubit comprises
additional Josephson junctions, by generalizing
\eref{wif_gen} as previously described
one can account for the quasiparticle losses in them. Unlike
Eq.~(\ref{wif_fq}) those losses remain finite at $\Phi_e=\Phi_0/2$, but can
be made small in a chain of $N \gg 1$ junctions of larger $E_J$,
as they scale with $1/N$ due to small phase fluctuations in the chain (see also~\cite{effind}).

In summary, we have presented a general approach to study
the quasiparticle relaxation mechanism
of superconducting qubits. It enables us to determine how
the qubit decay rate depends on the magnetic flux used to tune the system properties.
The method is applicable to
any superconducting qubit and arbitrary quasiparticle population.
For small phase fluctuations and transitions between neighboring levels, the
decay rate can be expressed in
terms of the real part of the ``classical''~\cite{BP} junction admittance, see \eref{wnn}, while
the imaginary part of the admittance determines the shift in the qubit frequency [\eref{freq_shift}].
This limit is applicable, {\sl e.g.}, to a single-junction transmon.
Decay rate and frequency shift have distinct dependencies on flux and quasiparticle distribution, so
comparing the two quantities may give information on the (possibly) nonequilibrium state of quasiparticles.

This work was supported by IARPA (ARO Contract No. W911NF-09-1-0369),
NSF (DMR Grant No. 0906498), and Yale University.
L.F. acknowledges partial support from CNR-Istituto di Cibernetica.


\begin{thebibliography}{50}
\bibitem{effind}This treatment is applicable to arrays of $N\gg 1$ junctions with small
phase fluctuations~\cite{flux_exp}.
The Hamiltonian of a qubit with few junctions of similar parameters is the sum of each
junction's charging and Josephson energies~\cite{orlando}. Then
\esref{HT} and \rref{wif_gen} are modified as explained in the paragraph preceding the one
containing \eref{vp0_Pe}.
\bibitem{negom}The expression for $\omega<0$
is obtained by the replacements $x \to x -\hbar\omega/\Delta$, $\omega \to -\omega$ in
the integrand.
\bibitem{footnote3}Therefore everywhere below we dispense with the
  probability of ``upward transitions'' for which $\omega_{if}<0$.
\bibitem{footnote}The flux dependence was not considered in Ref.~\cite{Martinis} --
there, the perturbation
driving the quasiparticle transitions was assumed to be independent of flux.
\bibitem{footnote2}At $E_J > E_L$ this ratio is not small and the
potential has multiple minima at some values of flux. Here we assume
non-degenerate wells -- see the discussion leading to \eref{wif_fq} for the
nearly-symmetric case.
\end{thebibliography}
\end{document}